\documentclass[aps,prd,twocolumn,groupedaddress]{revtex4}
\usepackage{exscale,latexsym,amsmath,amssymb} 
\usepackage[dvips]{graphicx}
\begin{document}
\title{Leading quantum gravitational corrections to scalar QED}
\author{N.~E.~J.~Bjerrum-Bohr}
\email[]{bjbohr@nbi.dk} 
\affiliation{The Niels Bohr Institute\\
Blegdamsvej 17, DK-2100 Copenhagen\\ Denmark}
\date{\today}
\begin{abstract}
We consider the leading post-Newtonian and quantum corrections to the non-relativistic
scattering amplitude of charged scalars in the combined theory of general relativity and
scalar QED.
The combined theory is treated as an effective field theory. This allows for a consistent
quantization of the gravitational field. 
The appropriate vertex rules are extracted from the action, and the non-analytic contributions
to the 1-loop scattering matrix are calculated in the non-relativistic limit. 
The non-analytical parts of the scattering amplitude, which are known to give the long range, low energy, 
leading quantum corrections, are used to construct the leading post-Newtonian and quantum corrections to the 
two-particle non-relativistic scattering matrix potential for two charged scalars. 
The result is discussed in relation to experimental verifications.   
\end{abstract}\maketitle
\section{Introduction}
In this paper we will treat general relativity as an effective quantum field theory.

The essential topic~\cite{Weinberg:1978kz,Donoghue:1993eb,Donoghue:dn}, 
in the concept of effective field theories is that the field couplings included in a certain 
Lagrangian are perturbatively determined by the energy scale of the problem, and not by the strict 
renormalization conditions one normally imposes in a quantum field theory. 
The effective action includes all terms consistent with the underlying 
symmetries of the theory. So in principle the effective action has an infinite number of terms.
Seen perturbatively the various terms of the action pertain to different energy scales of the theory so only 
a finite number of terms need to be taken into account at each loop order.
The Lagrangian is in some sense believed to be somewhat less fundamental than in normal renormalizable theories 
and has to be replaced by a more fundamental theory at sufficiently high energies --- but at low energies the 
effective Lagrangian presents an interesting path to avoid the traditional renormalization problems
of non-renormalizable theories.

As the action of an effective field theory includes all terms, any occurring field singularity of the 
theory will already correspond to certain terms of the action, and it will hence be possible to absorb 
such singularities into the coupling constants of the effective Lagrangian. Thus treating all
coupling constants as experimentally determined quantities, the effective field theory is finite at each loop order.

The old way of thinking of renormalizability of a field theory is not an issue when considering effective
field theories.  

It is well known that quantum field theories of pure general relativity,
as well as quantum theories for general relativity including scalar~\cite{Veltman1,Veltman2,Goroff:1985th},
fermion or photon fields~\cite{Deser:cz,Deser:cy}, suffer from severe problems with renormalizability in the 
traditional meaning of the word. 

A solution to this apparent obstacle is thus to treat general relativity as an effective 
field theory. The gravitational action then consists not solely of
the Einstein curvature term plus the minimal couplings of the matter terms, but 
of all terms consistent with general coordinate invariance of the theory. 
As an effective field theory, general relativity thus can be dealt with as any other 
quantum field theory.

General relativity with additional derivative couplings has been discussed
in the literature~\cite{Pais:1950za,Stelle:1976gc,Stelle:1977ry,Simon:ic}, and various
issues concerning general relativity as a classical or a quantum theory with higher 
derivative terms has been dealt with. 
However, treating general relativity as an effective field theory to find the leading pure 1-loop
gravitational corrections to the Newtonian potential was first done in~\cite{Donoghue:1993eb,Donoghue:dn}. 
Recently, some work has been carried out in~\cite{Donoghue:2001qc} using the same
technique to calculate the quantum corrections to the Reissner-Nordstr\"om and Kerr-Newman
metrics. We will discuss their result in relation to the potential.

For the non-analytical terms of a certain diagram to 1-loop order all vertex rules 
are given only by the Einstein curvature term plus the minimal coupling matter terms --- but
as we move on to higher loop calculations one will have to include the effects of
higher loop contributions too.
In this paper we will consider 1-loop effects and the lowest order theory is therefore
adequate for our purpose. 

We will extract the non-analytical parts of the full set of 1-loop diagrams needed for the
1-loop scattering matrix in the combined quantum theory of general relativity and scalar QED. 
As we shall see, the non-analytical contributions correspond to the long range corrections of the
potential.

We will employ the background field method first introduced in~\cite{Dewitt}. 
Here the gravitational background fields are not flat and the quantum corrections are added 
to the gravitational background field.

When nothing else is stated, we work with units $c=\hbar=1$ and employ the metric convention 
($1,-1,-1,-1$). 

The structure of the paper will be as follows. First we will review general relativity
as an effective field theory in more detail and see how to combine scalar QED with
general relativity.

Then we will look at the calculations of the diagrams. 
Finally we will construct the potential and discuss the result in relation to 
~\cite{Donoghue:2001qc,Barker:bx,Barker:ae,Iwasaki,Okamura}. The vertex rules 
are presented in appendix.  

\section{General relativity and scalar QED as a combined effective field theory}
A general covariant version of the scalar QED Lagrangian is
\begin{multline}
{\cal L} = \sqrt{-g}\bigg[-\frac14\left(g^{\alpha\mu}g^{\beta\nu}F_{\alpha\nu}F_{\mu\beta}\right) \\ + (D_\mu\phi+ieA_\mu)^*(g^{\mu\nu})(D_\nu\phi+ieA_\nu) - m^2|\phi|^2
\bigg]
\end{multline}
where $F_{\mu\nu} \equiv  {D}_\mu A_\nu - {D}_\nu A_\mu= \partial_\mu A_\nu - \partial_\nu A_\mu$, and ${D}_\mu$ denotes the 
covariant derivative with respect to the gravitational field, $g_{\mu\nu}$. As $\phi$ is a scalar ${D}_\mu \phi = \partial_\mu \phi$.  

We expand the effective Lagrangian in orders of magnitude of the derivative contributions. Derivatives of light fields $\tilde\partial$ will essentially go as powers of momentum, 
while derivatives of massive fields $\partial$ will generate powers of the interacting masses. As the interacting masses are often orders of magnitude higher
than the momentum terms --- the derivatives on the massive fields will often generate the leading contributions.
 
Counting the number of derivatives in each term of the above Lagrangian we see 
that the term with $g^{\alpha\mu}g^{\beta\nu}F_{\alpha\nu}F_{\mu\beta}$ goes as $\sim \tilde\partial\tilde\partial$,
while the scalar field terms goes as $\sim \partial\partial$ and $\sim 1$ respectively. 
Thus seen in the light of effective field theory, the above Lagrangian represents the minimal
derivative couplings of the gravitational fields to the photon and complex scalar fields.  

Typical 1-loop field singularities for the mixed graviton and photon fields in the minimal theory are known to take the form~\cite{Deser:cz}
\begin{eqnarray}
\sqrt{-g}T_{\mu\nu}^2, && \sqrt{-g}R_{\mu\nu}T^{\mu\nu}
\end{eqnarray}
where $T_{\mu\nu} = F_{\mu\alpha}F_{\nu}^{\ \alpha} -\frac14g_{\mu\nu}F^{\alpha\beta}F_{\alpha\beta}$ is the 
Maxwell stress tensor, and where $R^\mu_{\ \nu\alpha\beta} \equiv \partial_\alpha \Gamma^\mu_{\nu\beta}-
\partial_\beta \Gamma^\mu_{\nu\alpha}+\Gamma^\mu_{\sigma\alpha}\Gamma^\sigma_{\nu\beta}-\Gamma^\mu_{\sigma\beta}\Gamma^\sigma_{\nu\alpha}$ 
is the Einstein curvature tensor. Examples of similar 1-loop divergences for the mixed graviton and scalar fields are
\begin{eqnarray}\sqrt{-g}R^{\mu\nu} \partial_\mu \phi^* \partial_\nu \phi, & \sqrt{-g} R |\partial_\mu \phi|^2,
& \sqrt{-g} R m^2 |\phi|^2
\end{eqnarray}
The two photon contributions are seen to go as $\sim \tilde\partial\tilde\partial\tilde\partial\tilde\partial$, 
while the scalar contributions goes as $\sim \tilde\partial\tilde\partial\partial\partial$ and $\tilde\partial\tilde\partial$ respectively. 
So clearly the 1-loop singularities correspond to higher derivative couplings of the fields.

Of course there will also be examples of mixed terms with both photon, graviton and complex scalar fields. We will not consider any of these terms 
explicitly.  

As we calculate the 1-loop diagrams using the minimal theory, singular terms with higher derivative couplings of the fields will thus unavoidably appear. 
We however do no need to worry about these singularities explicitly, because the combined theory is treated as an effective field theory.

In order to treat the combined theory as an effective field theory we will include into the minimal derivative coupled Lagrangian a piece like,
\begin{equation}
{\cal L}_{\text{photon}} = \sqrt{-g}\Big[c_1 T_{\mu\nu}^2 + c_2R_{\mu\nu}T^{\mu\nu} + \ldots \Big]
\end{equation}
for the photon field and 
\begin{equation}\begin{split}
{\cal L}_{\text{scalar}} &= \sqrt{-g} \Big[d_1 R^{\mu\nu} \partial_\mu \phi^* \partial_\nu \phi + d_2 R |\partial_\mu \phi|^2\\&
+ d_3 R m^2 |\phi|^2 + \ldots \Big]
\end{split}\end{equation}
for the scalar field. The ellipses symbolize other higher derivative couplings at 1-loop order which are not included in the 
above equations, e.g. other higher derivative couplings and mixed contributions with both photon, graviton and scalar couplings.

The coefficients $c_1$, $c_2$, $d_1$, $d_2$ and $d_3, \ldots$ in the above equation are in the effective theory seen as energy 
scale dependent couplings constants to be measured experimentally. Every singular field term from the lowest order Lagrangian is thus 
absorbed into effective action, leaving us with a finite theory at 1-loop order, with a number of coupling coefficients to be
determined by experiment.
  
The effective combined theory of scalar QED and general relativity is thus in some sense a traditional 
renormalizable theory at 1-loop order. At low energies the theory is determined 
only by the minimal derivative coupled Lagrangian, however at very high energies, higher derivative terms will manifest themselves 
in measurable effects, and the unknown coefficients $c_1$, $c_2, \ldots$,  $d_1$, $d_2$ and $d_3, \ldots$ will have to be determined 
explicitly by experiment. 
This process of absorbing generated singular field terms into the effective action will of course have to continue at every loop order.

In order to compute the leading long range, low energy quantum corrections to this theory, it is useful to make a distinction between 
non-analytical and analytical contributions from the diagrams. The non-analytical contributions are inherently non-local effects which
cannot be expanded in a power series in momentum. The non-analytical effects comes from the propagation of massless particle modes 
such as gravitons and photons. 
This destinction originates from the impossibility of expanding a massless propagator $\sim \frac{1}{q^2}$ while 
we have the well known
\begin{equation}
\frac{1}{q^2-m^2} = -\frac{1}{m^2}\Big(1+\frac{q^2}{m^2}+\ldots \Big)
\end{equation}
expansion for the massive propagator. As seen no $\sim \frac{1}{q^2}$ terms is generated by the above expansion of the
massive propagator.
 
As we will see explicitly, these non-analytic contributions will be governed to leading order only by
the minimally coupled Lagrangian. 

The analytical contributions in the diagrams are local effects, which are always expandable in power series solutions. 

Typical examples of the non-analytical effects are e.g. terms which in the S-matrix go as 
$\sim \ln(-q^2)$ or $\sim\frac1{\sqrt{-q^2}}$, while the general example of an analytical effect is a power series in the momentum $q$. 
As we are only interested in non-local effects, we will only consider the non-analytical contributions.     

The high energy renormalization of the theory is thus of no concern for us --- as we are only finding the leading finite non-analytical 
momentum contributions for the 1-loop diagrams in the low energy scale of the theory. 
Hence the singular analytical momentum parts which have to be absorbed into coefficients of the higher derivative couplings, are of no
interest to us here and will not be manifested in this energy regime of the theory.   

We can now proceed with our quantization of general relativity and scalar QED as an effective field theory.

The quantization procedure will be as follows. 
We will define the metric as the sum of a background part $\bar g_{\mu\nu}$ and a quantum contribution $\kappa h_{\mu\nu}$, where $\kappa^2 = 32G\pi$
\begin{equation}
g_{\mu\nu}\equiv \bar g_{\mu\nu} + \kappa h_{\mu\nu}
\end{equation}
From this equation we get the expansions for the upper metric field $g^{\mu\nu}$ (defined to be the inverse matrix), and for $\sqrt{-g}$ (where $\det(g_{\mu\nu})=g$) as
\begin{equation}\begin{split}
g^{\mu\nu} &= \bar g^{\mu\nu} - \kappa h^{\mu\nu} + \ldots\\
\sqrt{-g} &= \sqrt{-\bar g}\left[1+ \frac12\kappa h + \ldots\right]
\end{split}\end{equation}
where $h^{\mu\nu}\equiv \bar g^{\mu\alpha} \bar g^{\nu\beta} h_{\alpha\beta}$ and $h = \bar g^{\mu\nu} h_{\mu\nu}$. We have only expanded to first order
in $h_{\mu\nu}$, as we need diagrams to second order in $\kappa$. 

\begin{widetext}
Next we expand the above covariant version of the scalar QED Lagrangian in terms of the fields. 
The result for the photon parts reads
\begin{equation}
{\cal L} = -\frac14\kappa h\left(\partial_\mu A_\alpha \partial^\mu A^\alpha - \partial_\mu
A_\alpha \partial^\alpha A^\mu\right) + \frac12\kappa h^{\mu\nu}\left(\partial_\mu A_\alpha
\partial_\nu A^\alpha + \partial_\alpha A_\mu \partial^\alpha A_\nu
- \partial_\alpha A_\mu \partial_\nu A^\alpha -\partial_\alpha A_\nu \partial_\mu 
A^\alpha\right)
\end{equation}
while the complex scalar part can be quoted as 
\begin{multline}
{\cal L} = \frac12\kappa h\left(|\partial_\mu\phi|^2 - m^2|\phi|^2\right) -\kappa h^{\mu\nu}(\partial_\mu\phi^*\partial_\nu\phi)
+\left(ieA_\mu\partial^\mu \phi^*\phi - ieA_\mu \phi^*\partial^\mu \phi\right)
+e^2A_\mu A^\mu |\phi|^2\\  + \frac12\kappa h\partial_\mu \phi^*(ieA^\mu)\phi-\frac12\kappa h(ieA^\mu)
\phi^*\partial_\mu \phi - \kappa h^{\mu\nu}\partial_\mu\phi^*(ieA_\nu)\phi + \kappa h^{\mu\nu}
(ieA_\mu)\phi^*\partial_\nu\phi
\end{multline}
\end{widetext}
From these equations one can find the vertex rules for the lowest order interaction
vertices of photons, complex scalars and gravitons for this theory. In the appendix 
we will present a summary of the vertex rules. 

\section{The results for the Feynman diagrams}
Before we consider the actual calculations of the diagrams we will take a
look on the general form for the scattering matrix.

The general form for diagrams contributing to the scattering matrix is 
\begin{equation}\begin{split}
{\cal M} &\sim \Big(A+B q^2 + \ldots + (\alpha_1 \kappa^2 + \alpha_2 e^2) \frac{1}{q^2} \\
&+\beta_1 e^2\kappa^2\ln(-q^2) + \beta_2 e^2\kappa^2 \frac{m}{\sqrt{-q^2}} + \ldots\Big)
\end{split}\end{equation}
where $A, B, \ldots$ correspond to the local analytical interactions and 
$\alpha_1, \alpha_2$ and $\beta_1, \beta_2, \ldots$ correspond to the leading non-analytical, non-local, 
long range interactions. 

The space parts of the non-analytical terms Fourier transform as,
\begin{equation}\begin{split}
\int \frac {d^3 k}{(2\pi)^3} \frac 1{|{\bf k}|^2}e^{i\bf{k}\cdot\bf{r}}& = \frac {1}{4\pi r}\\
\int \frac {d^3 k}{(2\pi)^3} \frac 1{|{\bf k}|}e^{i\bf{k}\cdot\bf{r}}& = \frac {1}{2\pi^2 r^2}\\
\int \frac {d^3 k}{(2\pi)^3} \ln({\bf k}^2)e^{i\bf{k}\cdot\bf{r}}& = \frac {-1}{2\pi r^3}\\
\end{split}\end{equation}
so clearly these terms will contribute to the long range corrections.

The non-analytical contribution, corresponding to the $\frac{1}{q^2}$ part, gives as seen the Newtonian and Coulomb potentials
respectively. The other non-analytical contributions generate the leading quantum and classical corrections to the 
Coulomb and Newtonian potentials in powers of $\frac1{r}$. It is necessary to have non-analytic contributions in the matrix element, to ensure that the S-matrix is unitary. 

The analytic contributions will not be considered in this work. As noted previously these corrections correspond to local interactions, and are
thus only needed for the high energy manifestation of the theory. Many of the analytical corrections will be divergent, and hence have to be carefully
absorbed into the appropriate terms of the effective action of the theory. 

We will not consider the radiative corrections due to soft bremsstrahlung in this approach. 
In some of the diagrams of this theory as well as in QED,
there are a need for introducing soft bremsstrahlung radiative corrections to the sum of the diagrams constituting the vertex corrections. 
We will not consider this aspect of the theory in this approach, as we are not computing the full amplitude of the S-matrix. Furthermore
certain effects have been included in the recent work of~\cite{Donoghue:2001qc}, where the gravitational vertex corrections are treated. 
This issue should be dealt with at some stage refining this effective theory of general relativity and scalar QED, however for now we will
carry on and simply compute the leading post-Newtonian and quantum corrections to the scattering, and leave this concern for future further investigations. 

\subsubsection{The definition of the potential}
The various definitions of the potential have been discussed at length in the literature. We will here define the potential
directly from the scattering matrix amplitude. 

In the quantization of general relativity the definition 
of the potential is certainly not obvious. One can choose between several definitions of the potential
depending on e.g. the physical situation, how to define the energy of the fields, the diagrams included etc.  

Clearly a valid choice of potential should be gauge invariant to be physically reasonable, 
but while other gauge theories like QCD allows a gauge invariant Wilson loop definition --- e.g. for a
quark-anti-quark potential, this is not directly possible in general relativity. 

There has however been some attempts to make a Wilson loop equivalent potential for quantum gravity. A
Wilson-like potential seems to be possible to construct in general relativity using the Arnowitt-Deser-Misner formula for the total
energy of the system~\cite{Modanese:1994bk}. 
This choice has been discussed in~\cite{Muzinich:1995uj} in the case of pure gravity coupled to scalar fields.   

A recent suggestion~\cite{Kazakov:2000mu} is that one should look at the full set of diagrams constituting the 1-loop scattering
matrix, and use the total sum of the 1-loop diagrams to decide the non-relativistic potential. As the full 1-loop
scattering matrix is involved, this choice of potential gives a gauge invariant definition.

This choice of potential is equivalent to that of~\cite{Hamber:1995cq}, where the scalar source pure gravity potential 
was treated. 

This choice of potential which includes all 1-loop diagrams seems to be the simplest, gauge invariant choice one
can make for the potential.       

We will calculate the non-relativistic potential using the the full amplitude. Here we simply relate the expectation
value for the $iT$ matrix to the Fourier transform of the potential $\tilde V({\bf q})$ in 
the non-relativistic limit as
\begin{equation}
\langle k_1,k_2 | i T | k_1', k_2' \rangle = -i\tilde {V}({\bf q})(2\pi)\delta(E - E') 
\end{equation}
where $k_1$, $k_2$ and $k_1'$, $k_2'$ are the incoming and outgoing momentum respectively, and $E-E'$ are the energy  
difference between the incoming and outgoing states~\cite{Iwasaki2}. Comparing this to the definition of the invariant matrix element
$i{\cal M}$ we get from the diagrams 
\begin{equation}
\langle k_1,k_2 | i T | k_1', k_2' \rangle = (2\pi)^4\delta^{(4)}(k_1-k_1'+k_2-k_2')(i{\cal M}) 
\end{equation}
we see that (we have divided the above equation with $2m_1 2m_2$ to obtain the 
non-relativistic limit)
\begin{equation}
\tilde V({\bf q}) = -\frac{1}{2m_1}\frac{1}{2m_2}{\cal M}
\end{equation}
so that 
\begin{equation}
V({\bf x}) = -\frac{1}{2m_1}\frac{1}{2m_2}\int \frac{d^3k}{(2\pi)^3}e^{i{\bf k} \cdot {\bf x}}{\cal M} 
\end{equation}
This will be our definition of the non-relativistic potential generated by the considered non-analytic parts, 
where ${\cal M}$ is the non-analytical part of the amplitude of the scattering process to a given loop order. This 
definition of the potential is also used in~\cite{Hamber:1995cq}.

\subsection{The diagrams contribution to the non-analytical parts of the scattering matrix}
Of the diagrams contributing to the scattering matrix only a certain class of diagrams
will actually contribute to the sum of non-analytical terms considered here --- the logarithmic
and square-root parts. In this treatment we will only consider the diagrams which contribute 
with non-analytical contributions in detail. Diagrams with many massive propagators will usually 
only contribute with analytical terms. Some of the diagrams have a somewhat complicated algebraic structure 
due to the involved vertex rules. To do the diagrams we developed an algebraic program for Maple 
(Waterloo software). 
Our program contract the various indices and performs the loop integrations.
In the following we will go through the diagrams and discuss how they are calculated in detail. We will begin 
with the tree diagrams.  

\subsubsection{The tree diagrams}
\begin{figure}[h]\vspace{0.5cm}
\begin{minipage}{0.43\linewidth}
\begin{center}
\includegraphics[scale=1]{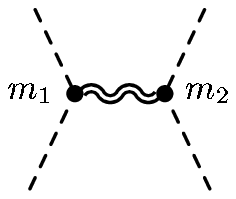}
\end{center}
{\centering 1(a)}
\end{minipage}
\begin{minipage}{0.43\linewidth}
\begin{center}
\includegraphics[scale=1]{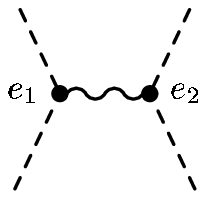}
\end{center}
{\centering 1(b)}
\end{minipage}
\caption{The set of tree diagrams contributing to the potential.}
\label{tree}\end{figure}

The set of tree diagrams contributing to the scattering matrix are those of figure \ref{tree}.
The formal expression for these diagrams are
\begin{equation}
{i \cal M}_{\rm 1(a)} = \tau_2^{\mu\nu}(k_1,k_2,m_1)\bigg[ \frac{i{\cal P}_{\mu\nu\alpha\beta}}{q^2}\bigg] \tau_2^{\alpha\beta}(k_3,k_4,m_2)  
\end{equation}
and 
\begin{equation}
{i \cal M}_{\rm 1(b)} = \tau_1^{\mu}(k_1,k_2,e_1)\bigg[ \frac{-i{\eta}_{\mu\nu}}{q^2}\bigg] \tau_1^{\nu}(k_3,k_4,e_2)  
\end{equation}
These diagrams yield no complications.
Contracting all indices and preforming the Fourier transforms one ends up with
\begin{equation}
V_{\rm 1(a)}(r) = -\frac{G m_1 m_2}{r}
\end{equation}
\begin{equation}
V_{\rm 1(b)}(r) = \frac{e_1 e_2}{4\pi r} 
\end{equation}
where $e_1,m_1$ and $e_2,m_2$ are the two
charges and masses of the system respectively. 
This is of course the expected results for these
diagrams. One gets the Newtonian and Coulomb 
terms for the potential of two charged scalars. 

The next class of diagrams we will consider is that of box diagrams. 
\subsubsection{The box diagrams and crossed box diagrams }
\begin{figure}[h]\vspace{0.5cm}
\begin{minipage}{0.43\linewidth}
\begin{center}
\includegraphics[scale=1.15]{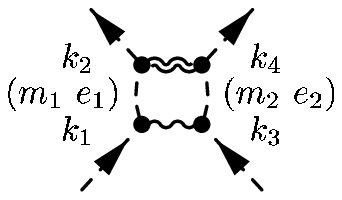}
\end{center}
{\centering 2(a)}
\end{minipage}
\begin{minipage}{0.43\linewidth}
\begin{center}
\includegraphics[scale=1.15]{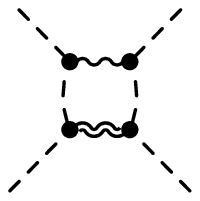}
\end{center}
{\centering 2(b)}
\end{minipage}\vspace{0.5cm}
\begin{minipage}{0.43\linewidth}
\begin{center}
\includegraphics[scale=1.15]{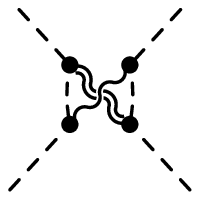}
\end{center}
{\centering 2(c)}
\end{minipage}
\begin{minipage}{0.43\linewidth}
\begin{center}
\includegraphics[scale=1.15]{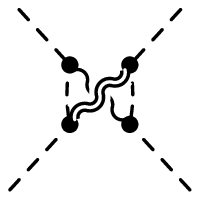}
\end{center}{\centering 2(d)}
\end{minipage}
\caption{The set of box and crossed box diagrams contributing to the non-analytical parts of the potential.}\label{box}
\end{figure}
There are four distinct diagrams. See figure \ref{box}. Two crossed box and two box diagrams.  
We will not treat all diagrams separately but rather discuss one of the
diagrams in detail and then present the total result for the diagrams. 
The diagram 2(a) is defined in the following way
\begin{equation}\begin{split}
{i \cal M}_{\rm 2(a)} & = \int\frac{d^4l}{(2\pi)^4}\frac{i}{(l+k_1)^2-m_1^2}\frac{i}{(l-k_3)^2-m_2^2}\\&
\times \tau_1^{\gamma}(k_1,k_1+l,e_1)\bigg[ \frac{-i\eta_{\gamma\delta}}{l^2}\bigg] \tau_1^{\delta}(k_3,-l+k_3,e_2)\\& 
\times \tau_2^{\mu\nu}(l+k_1,k_2,m_1)\bigg[ \frac{i{\cal P}_{\mu\nu\sigma\rho}}{(l+q)^2}\bigg] \tau_2^{\sigma\rho}(k_3-l,k_4,m_2)
\end{split}\end{equation}
where we have chosen a certain parameterization of the momenta in the diagram, 
the side with mass ($m_1$) and charge ($e_1$) has $k_1$, $k_2$ as incoming and outgoing momentum respectively. Correspondingly
the other side with mass ($m_2$) and charge ($e_2$) has $k_3$, $k_4$ as incoming and outgoing momentum respectively

The algebraic structure of this diagrams is rather involved and complicated, but
yields no complications using our algebraic program. The integrals are rather 
complicated to do, but one can make use of various contraction rules for the integrals which holds true on the mass-shell. 

From the choice $q=k_1-k_2=k_4-k_3$ one can easily derive:
\begin{equation}\begin{split}
k_1\cdot q & = k_4 \cdot q = -k_2\cdot q = -k_3 \cdot q = \frac{q^2}{2}\\
k_1\cdot k_2 & = m_1^2 - \frac{q^2}{2}\\
k_3\cdot k_4 & = m_2^2 - \frac{q^2}{2}
\end{split}\end{equation}
where $k_1^2=k_2^2=m_1^2$ and  $k_3^2=k_4^2=m_2^2$ on the mass shell.

On the mass shell we have identities like:
\begin{equation}\begin{split}
l \cdot q & = \frac{(l+q)^2-q^2-l^2}2\\
l \cdot k_1 & = \frac{(l+k_1)^2-m_1^2-l^2}2\\
l \cdot k_3 & = -\frac{(l-k_3)^2-m_2^2-l^2}2
\end{split}\end{equation}
\begin{widetext}
Now clearly e.g.:
\begin{equation}\begin{split}
\int\frac{d^4l}{(2\pi)^4} \frac{l\cdot q} {l^2(l+q)^2((l+k_1)^2-m_1^2)((l-k_3)^2-m_2^2)}  = 
\frac12\int\frac{d^4l}{(2\pi)^4} \frac{(l+q)^2-q^2-l^2} {l^2(l+q)^2((l+k_1)^2-m_1^2)((l-k_3)^2-m_2^2)} 
\end{split}\end{equation}
as only the integral with $q^2$ yield the non-analytical terms we let
\begin{equation}
\int\frac{d^4l}{(2\pi)^4} \frac{l\cdot q} {l^2(l+q)^2((l+k_1)^2-m_1^2)((l-k_3)^2-m_2^2)} \rightarrow 
\frac{-q^2}{2}\int\frac{d^4l}{(2\pi)^4} \frac{1} {l^2(l+q)^2((l+k_1)^2-m_1^2)((l-k_3)^2-m_2^2)} 
\end{equation}
A perhaps more significant reduction of the integrals is with the contraction of the sources momenta, e.g.
 \begin{equation}\begin{split}
\int\frac{d^4l}{(2\pi)^4} \frac{l\cdot k_1} {l^2(l+q)^2((l+k_1)^2-m_1^2)((l-k_3)^2-m_2^2)} & =  
\frac12\int\frac{d^4l}{(2\pi)^4} \frac{(l+k_1)^2-m_1^2-l^2} {l^2(l+q)^2((l+k_1)^2-m_1^2)((l-k_3)^2-m_2^2)} \\ & 
\rightarrow 
\frac12\int\frac{d^4l}{(2\pi)^4} \frac{1}{l^2(l+q)^2((l-k_3)^2-m_2^2)} 
\end{split}\end{equation}
or 
\begin{equation}\begin{split}
\int\frac{d^4l}{(2\pi)^4} \frac{l\cdot k_3} {l^2(l+q)^2((l+k_1)^2-m_1^2)((l-k_3)^2-m_2^2)} = & 
\int\frac{d^4l}{(2\pi)^4} \frac{-(l-k_3)^2+m_2^2+l^2} {l^2(l+q)^2((l+k_1)^2-m_1^2)((l-k_3)^2-m_2^2)} 
\\ & \rightarrow 
-\frac12\int\frac{d^4l}{(2\pi)^4} \frac{1} {l^2(l+q)^2((l+k_1)^2-m_1^2)} 
\end{split}\end{equation}
as seen the contraction of a loop momentum factor with a sources momentum factor removes one
of the propagators leaving a much simpler loop integral.  \end{widetext}

Such reductions in the box diagram integrals help to do the calculations. The remaining integrals can be
done quite easily and results are presented in the appendix, together with the lowest box integral which
has to be done explicitly. 

The final sum for these diagrams gives
\begin{equation}
V_{\rm 2(a)+2(b)+2(c)+2(d)}(r) = \frac{10Ge_1e_2}{3\pi^2 r^3} 
\end{equation}

\subsubsection{The triangular diagrams}
\begin{figure}[h]
\begin{minipage}{0.43\linewidth}
\begin{center}
\includegraphics[scale=1.1]{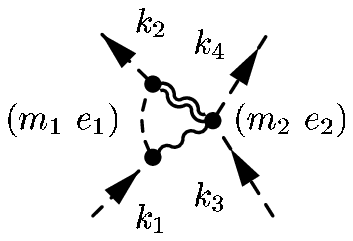}
\end{center}
{3(a)}
\end{minipage}
\begin{minipage}{0.43\linewidth}\vspace{0.4cm}
\begin{center}
\includegraphics[scale=1.1]{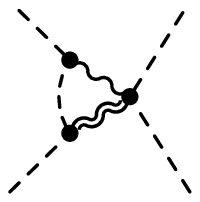}
\end{center}
{\vspace{0.15cm}3(b)}
\end{minipage}\vspace{0.3cm}
\begin{minipage}{0.43\linewidth}
\begin{center}
\includegraphics[scale=1.1]{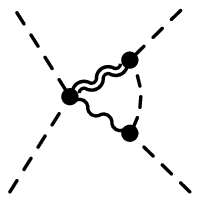}
\end{center}
{3(c)}
\end{minipage}
\begin{minipage}{0.43\linewidth}
\begin{center}
\includegraphics[scale=1.1]{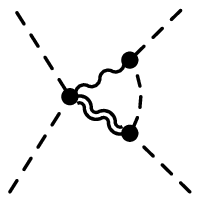}
\end{center}
{3(d)}
\end{minipage}
\caption{The set of triangular diagrams contributing to the non-analytical terms of the potential.}\label{trig}
\end{figure}

The following triangular diagrams contributes with non-analytic contributions to the potential. See figure \ref{trig}.
As for the box diagrams we will only consider one of the 
diagrams --- here again the first namely 3(a). The formal expression
for this particular diagram is --- we just apply the vertex rules:
\begin{equation}\begin{split}
{i \cal M}_{\rm 3(a)} & = \int\frac{d^4l}{(2\pi)^4}\frac{i}{(l+k_1)^2-m_1^2}\\&
\times \tau_1^{\gamma}(k_1,k_1+l,e_1)\bigg[ \frac{-i\eta_{\gamma\delta}}{l^2}\bigg] \tau_2^{\mu\nu}(l+k_1,k_2,m_1)\\ & 
\times \bigg[ \frac{i{\cal P}_{\mu\nu\sigma\rho}}{(l+q)^2}\bigg] \tau_5^{(\delta)\sigma\rho}(k_3,k_4,e_2)
\end{split}\end{equation}
Again all the needed integrals are of the type discussed in the appendix. Applying our contraction program and doing the integrations
leaves us with a result, which Fourier transformed yields the following contribution
to the potential
\begin{equation}
V_{\rm 3(a)+3(b)+3(c)+3(d)}(r) = \frac{Ge_1e_2(m_1+m_2)}{\pi r^2} - \frac{4e_1e_2G}{\pi^2 r^3}
\end{equation}
As seen these diagrams yield both a classical the $\sim \frac{1}{r^2}$ contribution, as well as an
quantum correction $\sim \frac{1}{r^3}$. 

\subsubsection{The circular diagram}
\begin{figure}[h]
\begin{center}
\includegraphics[scale=1.1]{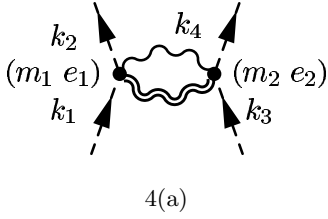}
\end{center}
{4(a)}
\caption{The circular diagram with non-analytic contributions.}\label{circ}
\end{figure}
The circular diagram, see figure \ref{circ}, has the following formal expression
\begin{equation}\begin{split}
{i \cal M}_{\rm 4(a)} & = \int\frac{d^4l}{(2\pi)^4} \tau_5^{\mu\nu(\gamma)}(k_1,k_2,e_1)\bigg[ \frac{-i\eta_{\gamma\delta}}{l^2}\bigg] \\& 
\times \bigg[ \frac{i{\cal P}_{\mu\nu\sigma\rho}}{(l+q)^2}\bigg] \tau_5^{\sigma\rho(\delta)}(k_3,k_4,e_2)
\end{split}\end{equation}
Doing the contractions and integrations gives the following contribution
to the potential
\begin{equation}
V_{\rm 4(a)}(r) = \frac{2Ge_1e_2}{\pi^2 r^3} 
\end{equation}

\subsubsection{1PR-diagrams}
\begin{figure}[h]
\begin{minipage}{0.43\linewidth}
\begin{center}
\includegraphics[scale=1.1]{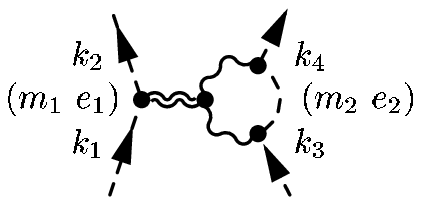}
\end{center}
{5(a)}
\end{minipage}
\begin{minipage}{0.1\linewidth}\ \end{minipage}
\begin{minipage}{0.43\linewidth}
\begin{center}
\includegraphics[scale=1.1]{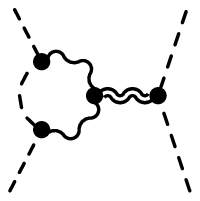}
\end{center}
{5(b)}
\end{minipage}\vspace{0.3cm}
\begin{minipage}{0.43\linewidth}
\begin{center}
\includegraphics[scale=1.1]{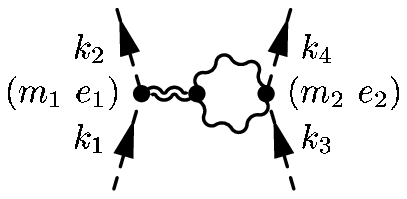}
\end{center}
{5(c)}
\end{minipage}
\begin{minipage}{0.1\linewidth}\ \end{minipage}
\begin{minipage}[scale=1.1]{0.43\linewidth}
\begin{center}
\includegraphics[scale=1.1]{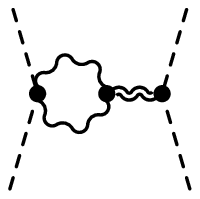}
\end{center}
{5(d)}
\end{minipage}
\caption{The class of the graviton 1PR vertex corrections which yield non-analytical corrections to the potential.}\label{gravi2n}
\end{figure}

The following class of the set of 1PR-diagrams corresponding to the gravitational
vertex correction will contribute to the potential, see figure \ref{gravi2n}.

Again we will not treat all diagrams separately. Instead we will consider two of the diagrams in details --- namely the diagrams 5(a)
and 5(c). First we will present the formal expressions for the diagrams using the vertex Feynman rules. Next we will briefly consider
the calculations and finally we will present the results. 

The formal expression for 5(a) is, 
\begin{equation}\begin{split}
{i \cal M}_{\rm 5(a)} & = \int\frac{d^4l}{(2\pi)^4}\frac{i}{(l-k_3)^2-m_2^2} \tau_2^{\mu\nu}(k_1,k_2,m_1)
\\& \times \bigg[ \frac{i{\cal P}_{\mu\nu\rho\sigma}}{q^2}\bigg] \tau_3^{\rho\sigma(\gamma\delta)}(l,l+q)
\tau_1^{\alpha}(k_3,k_3-l,e_2)\\ & \times\bigg[ \frac{-i\eta_{\alpha\gamma}}{l^2}\bigg]\bigg[ \frac{-i\eta_{\beta\delta}}{(l+q)^2}\bigg] \tau_1^{\beta}(k_3-l,k_4,e_2)
\end{split}\end{equation}
while the expression for 5(c) reads
\begin{equation}\begin{split}
{i \cal M}_{\rm 5(c)} & = \int\frac{d^4l}{(2\pi)^4}
\tau_2^{\mu\nu}(k_1,k_2,m_1) \bigg[ \frac{i{\cal P}_{\mu\nu\rho\sigma}}{q^2}\bigg]\\ & \times \tau_3^{\rho\sigma(\gamma\delta)}(l,l+q)
\tau_4^{\alpha\beta}(k_3,k_4,e_2)\bigg[ \frac{-i{\eta}_{\gamma\alpha}}{l^2}\bigg]\bigg[ \frac{-i{\eta}_{\delta\beta}}{(l+q)^2}\bigg]
\end{split}\end{equation}

Again the calculations of these diagrams diagrams yield no real complications using our algebraic program.

The result for the diagrams $5(a-d)$ are in terms of the 
corrections to the potential
\begin{equation}\begin{split}
V_{\rm 5(a)+5(b)+5(c)+5(d)}(r) & = \frac{G(e_2^2m_1+e_1^2m_2)}{8\pi r^2}\\ &  - \frac{G\big(\frac{m_1}{m_2}e_2^2+\frac{m_2}{m_1}e_1^2\big)}{3\pi^2 r^3}
\end{split}\end{equation}
where we have associated a factor of one-half due to the symmetry of the diagrams 5(c-d).

We have checked explicitly, that the above result for correction to the potential is in complete agreement, with the result for the 
gravitational vertex correction calculated in~\cite{Donoghue:2001qc}.     

\begin{figure}[h]
\begin{minipage}{0.43\linewidth}
\begin{center}
\includegraphics[scale=1.1]{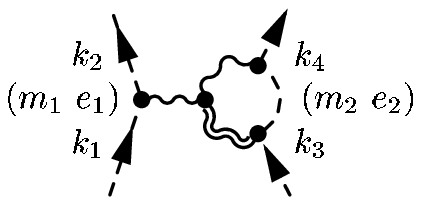}
\end{center}
{6(a)}
\end{minipage}
\begin{minipage}{0.1\linewidth}\ \end{minipage}
\begin{minipage}{0.43\linewidth}
\begin{center}
\includegraphics[scale=1.1]{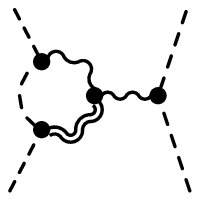}
\end{center}
{6(b)}
\end{minipage}\vspace{0.3cm}
\begin{minipage}{0.43\linewidth}
\begin{center}
\includegraphics[scale=1.1]{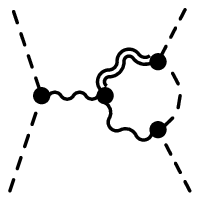}
\end{center}{6(c)}
\end{minipage}
\begin{minipage}{0.1\linewidth}\ \end{minipage}
\begin{minipage}{0.43\linewidth}
\begin{center}
\includegraphics[scale=1.1]{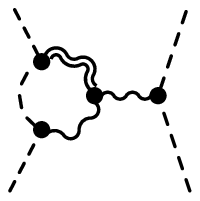}
\end{center}
{6(d)}
\end{minipage}
\caption{The first class of the photon vertex 1PR corrections which yield non-analytical corrections to the potential.}
\label{pho2n1}
\end{figure}

For the photonic vertex correction we consider the following diagrams. See figure \ref{pho2n1}.

\begin{figure}[h]
\begin{minipage}{0.45\linewidth}
\begin{center}
\includegraphics[scale=1.1]{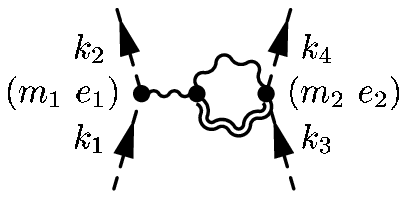}
\end{center}
{\centering 7(a)}
\end{minipage}
\begin{minipage}{0.1\linewidth}\ \end{minipage}
\begin{minipage}{0.35\linewidth}
\begin{center}
\includegraphics[scale=1.1]{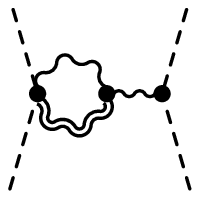}
\end{center}{\centering 7(b)}
\end{minipage}
\caption{The remaining photonic vertex 1PR diagrams which yield non-analytical corrections to the potential.}
\label{pho2n2}
\end{figure}

Together with the diagrams. See figure \ref{pho2n2}.

Again we look upon the formal expression for only two of the diagrams --- namely 6(a) and 7(a)
\begin{equation}\begin{split}
{i \cal M}_{\rm 6(a)} & = \int\frac{d^4l}{(2\pi)^4}\frac{i}{(l-k_3)^2-m_2^2}
\tau_1^{\gamma}(k_1,k_2,e_1)\\&\times\bigg[ \frac{-i\eta_{\gamma\delta}}{q^2}\bigg] \tau_3^{\sigma\rho(\delta\alpha)}(q,l+q)
\tau_2^{\mu\nu}(k_3,k_3-l,m_2)\\&\times\bigg[ \frac{i{\cal P}_{\mu\nu\sigma\rho}}{l^2}\bigg]\bigg[ \frac{-i{\eta}_{\beta\alpha}}{(l+q)^2}\bigg] \tau_1^{\beta}(k_3-l,k_4,e_2)
\end{split}\end{equation}
\begin{equation}\begin{split}
{i \cal M}_{\rm 7(a)} & = \int\frac{d^4l}{(2\pi)^4}
\tau_1^{\gamma}(k_1,k_2,e_1)\bigg[ \frac{-i\eta_{\gamma\delta}}{q^2}\bigg] \tau_3^{\mu\nu(\delta\alpha)}(q,l+q)\\& 
\times \bigg[ \frac{i{\cal P}_{\mu\nu\sigma\rho}}{l^2}\bigg]\bigg[ \frac{-i{\eta}_{\alpha\beta}}{(l+q)^2}\bigg] \tau_5^{\sigma\rho(\beta)}(k_3,k_4,e_2)
\end{split}\end{equation}

The result for the diagrams $6(a-d)+7(a-b)$ are in terms of the
corrections to the potential
\begin{equation}
V_{\rm 6(a)+6(b)+6(c)+6(d)+7(a)+7(b)}(r) = -\frac{Ge_1e_2(m_1+m_2)}{4\pi r^2}
\end{equation}

\subsubsection{The vacuum polarization diagram}
\begin{figure}[ht]
\begin{center}
\includegraphics[scale=1.1]{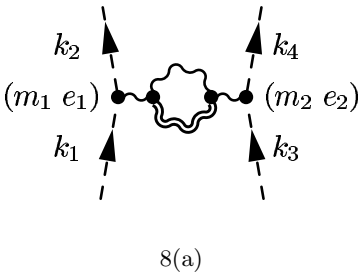}
\end{center}
{8(a)}
\caption{The only mixed vacuum polarization diagram to contribute to the potential. There is no mixed
corresponding ghost diagram associated with this diagram.}\label{vacuum}  
\end{figure}

The vacuum diagram, see figure \ref{vacuum}, has the following formal expression
\begin{equation}\begin{split}
{i \cal M}_{\rm 8(a)} & = \int\frac{d^4l}{(2\pi)^4}
\tau_1^{\gamma}(k_1,k_2,e_1)\bigg[ \frac{-i\eta_{\gamma\delta}}{q^2}\bigg]\tau_3^{\sigma\rho(\delta\alpha)}(q,-l)\\&\times
\tau_3^{\mu\nu(\beta\epsilon)}(-l,q)\bigg[ \frac{i{\cal P}_{\mu\nu\sigma\rho}}{(l+q)^2}\bigg]
\bigg[ \frac{-i{\eta}_{\beta\alpha}}{l^2}\bigg]\bigg[ \frac{-i{\eta}_{\epsilon\phi}}{q^2}\bigg]\\&\times \tau_1^{\phi}(k_3,k_4,e_2)
\end{split}\end{equation}

It gives the following contribution to the
potential
\begin{equation}
V_{\rm 8(a)}(r) = \frac{Ge_1e_2}{6 \pi^2 r^3} 
\end{equation}
 
The exact photon contributions for the 1-loop divergences of the minimal theory can be found in~\cite{Deser:cz}. 
Using that the pole singularity $\frac{1}\epsilon$ will always be followed by a $\ln(-q^2)$ contribution, one can 
read off the non-analytic result for the loop diagram using the coefficient of the singular pole term. We have explicitly 
checked our result for this diagram with the result derived in this fashion.

\begin{figure}
\begin{minipage}{0.25\linewidth}
\begin{center}
\includegraphics[scale=1]{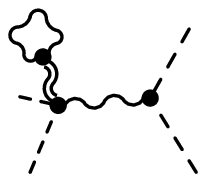}
\end{center}A
\end{minipage}
\begin{minipage}{0.03\linewidth}\ \end{minipage}
\begin{minipage}{0.25\linewidth}\begin{center}
\includegraphics[scale=1.1]{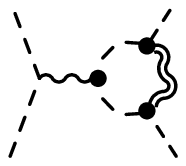}
\end{center}B
\end{minipage}
\begin{minipage}{0.03\linewidth}\ \end{minipage}
\begin{minipage}{0.25\linewidth}\begin{center}
\includegraphics[scale=1.02]{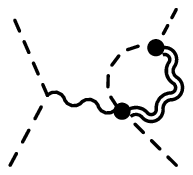}
\end{center}C
\end{minipage}
\caption{Diagrams which will only give contributions to the analytical parts of the potential.}\label{other}
\end{figure}

The above diagrams generate all the non-analytical contributions to the potential. There are other diagrams contributing to the 
1-loop scattering matrix, but those diagrams will only give analytical contributions, so we will not discuss them here in much detail. 
Examples of these diagrams are shown, see figure \ref{other}.
 
The diagram A is a tadpole. Tadpole diagrams will never depend on the transverse momentum of the diagrams, and will thus
never contribute with a non-analytical term. In fact, massless tadpoles will be zero in dimensional regularization. 
The diagram B is interesting --- as it is of the same type as the diagrams 6(a-d),
however with two massive propagators and one massless instead of two massless and one massive propagator. One can show that this diagram
will not contribute with non-analytic terms, because such an integral with two massive denominators and one massless will only give 
analytical contributions. In the case of diagram C, the loop is on one of the external legs. Hence the loop integrations will not
depend on the interchanged momentum of the diagram. Thus it cannot give any non-analytical contributions to the potential.

\section{The result for the potential}
Adding it all up, the final result for the potential reads
\begin{equation}\begin{split}
V(r) &= -\frac{Gm_1m_2}{r} + 
\frac{\tilde\alpha \tilde e_1 \tilde e_2}{r} +\\ &  
+\frac12 \frac{(m_1\tilde e_2^2+m_2\tilde e_1^2)G\tilde\alpha}{c^2r^2}
+\frac{3\tilde e_1\tilde e_2(m_1+m_2)G\tilde\alpha}{c^2r^2}\\ & 
-\frac43 \Big(\frac{m_1^2\tilde e_2^2+m_2^2\tilde e_1^2}{m_1m_2}\Big)\frac{G\tilde\alpha\hbar}{\pi c^3r^3}
+6 \frac{\tilde e_1\tilde e_2G\tilde\alpha\hbar}{\pi c^3r^3}
\end{split}\end{equation} 
including the appropriate physical factors of $\hbar$ and $c$, and rescaling everything
in terms of $\tilde \alpha = \frac{\hbar c}{137}$. The charges $\tilde e_1$
and $\tilde e_2$ are normalized in units of the elementary charge.

We see that there are various different types of terms in this expression. The first
two terms are, as noted before, the well-known Newtonian and Coulomb terms. They
represent the lowest order interactions of the two sources. These terms are as expected,
and they will dominate the potential at sufficiently low energies. 

The next two terms are the classical post-Newtonian corrections to the potential. These
terms are the leading post-Newtonian corrections, which are also present in general relativity with the 
inclusion of charged particles. A result for the post-Newtonian corrections derived from classical 
considerations can be found in~\cite{Barker:bx}, and in our notation reads:
\begin{equation}\begin{split}
V_{\rm post-Newtonian} &= \frac12 \frac{(m_1\tilde e_2^2+m_2\tilde e_1^2)G\tilde\alpha}{c^2r^2}\\&
+(\alpha_p+\alpha_g-1)\frac{\tilde e_1\tilde e_2(m_1+m_2)G\tilde\alpha}{c^2r^2}
\end{split}\end{equation}
We see that classical expectations for the post-Newtonian correction terms exactly match the ones we have derived, and 
that the coefficient of the first term is exactly equal to ours. 
The result for the second term is equivalent in form to the term we have derived, 
and the coefficient can be made to match our result exactly for particular values of $\alpha_p$ and $\alpha_g$. The physical
significance of the arbitrary parameters $\alpha_g$ and $\alpha_p$ however requires some explanation.
The values of $\alpha_p$ and $\alpha_g$ are coordinate dependent coefficients for the potential, which can take
arbitrary values depending on the coordinate system chosen to represent the potential. The  
non-relativistic potential is ambiguous in this sense~\cite{Barker:bx,Barker:ae}. This fact has also been discussed in 
~\cite{Iwasaki,Okamura} in the case of the pure gravitational potential.  

The parameter $\alpha_g$ is related to the gravitational propagator while the coefficient $\alpha_p$ is related to the photonic
propagator. In a forthcoming publication~\cite{grav}, we will consider the pure gravitational post-Newtonian corrections to the
potential of two scalars, and discuss the result in relation to the coefficient $\alpha_g$. 

An interesting observation is that the first term of the post-Newtonian correction is invariant under the coordinate transformation. 
This term originates as noted previously from the gravitational vertex corrections which generate the corrections to
the classical metric, see~\cite{Donoghue:2001qc}. This suggests that it may be better to consider corrections to a classical metric, than
to a non-relativistic potential. The post-Newtonian and quantum corrections to the Schwarzschild and Kerr metrics will be considered
in a forthcoming publication~\cite{metric}.  

The last two contributions are the most interesting from a quantum point of view. These
two terms represent the leading 1-loop quantum corrections to the mixed theory of
general relativity and scalar QED here computed for the first time. 
As seen in SI units $\frac{\hbar G}{c^3}\sim 10^{-70} \ {\rm meters}^2$, 
so these corrections are very small indeed, and hence seemingly impossible to detect
experimentally. This is especially due to the large terms of the Coulomb and Newtonian terms.

We have checked explicitly that the coordinate transformations of~\cite{Barker:bx} 
which affect the second post-Newtonian term, cannot alter the
coefficients of the quantum contributions to the corrections of the non-relativistic potential. 

When looking at the expression for the potential, one notices that the post-Newtonian and
the quantum correction to the potential are split up into two types of terms. 
There is one term where the two charges are multiplied together and one term where the two
charges are squared and separated. We have assumed that the particles are not identical. For identical
particles the two type of terms must be exactly identical in form. For identical particles one should include
the appropriate diagrams with crossed particle lines. 

When one of the particles is either very large or with a very high charge, some of the
contributions will dominate over others. 
The terms with separated charges will correspond to the dominating 
terms, if one of the scattered masses or charges were much larger than other. In this case
the gravitational vertex corrections will generate the dominating leading contribution to 
the potential. E.g.
with a very high charge for one of the particles --- the probing particle will fell the most
the gravitational effect coming from the electromagnetic field surrounding the
heavily charged particle. 

For a very large mass $(M \sim 10^{30}\ {\rm kg})$ but a very low charged particle 
$\tilde e_1 \sim 0$ (the Sun), and a charged $(\tilde e_2 \sim 1)$ but very low mass
particle $(m \sim 10^{-31}\ {\rm kg})$ (an electron), one could perhaps test this effect
experimentally, because then the Newton effect is $\frac{GmM}r \sim \frac{10^{-10} \ {\rm J \cdot meters}}{r}$, but the quantum effect 
$G\hbar\tilde\alpha \frac{\tilde e_2^2 M}{mc^3r^3} \sim \frac{10^{-37}\ {\rm J\cdot meters}^3}{r^3}$, while the classical contribution
$GM\frac{\tilde e_2^2\tilde\alpha}{c^2r^2}\sim \frac{10^{-25}\ {\rm J \cdot meters}^2}{r^2}$. 
The ratio between the post-Newtonian effects and the quantum correction is for this experimental
setup still very large, but not quite as impossible as often seen in quantum gravity.  

Experimental verifications of general relativity as an effective field will perhaps be a
very difficult task. The problem is caused by the normally very large classical
expectations of the theory. These expectations implies that nearly any quantum effect in powers of $G\hbar$
will be nearly neglectable compared to $G$. Therefore the quantum effects will be very hard
to extract, using measurements where classical expectations are involved. The solution to 
this obstacle could be to magnify a certain quantum effect. This could be in cases where
the classical effect where independent of the energy scale, but where the quantum effects were
largely effected by the energy scale. Such an effect would only be observed, when very large interaction
energies are involved. 

Another way to observe a quantum effect could be when a certain classical expectation is
zero, but the quantum effect would yield a contribution. A quantum gravitational ''anomaly''. Such 'null'
experiments maybe used to test a quantum theory for gravity~\cite{Donoghue2}.    

\section{Discussion}
Normally general relativity is viewed as a non-renormalizable theory, and consequently
a quantum theory for general relativity is believed to be an inconsistent theory. 
However, treated as an effective field theory, the renormalization inconsistency of general 
relativity is not an issue --- as the theory can be explicitly renormalized to any given
loop order. This fact was first explored in~\cite{Donoghue:1993eb,Donoghue:dn}. 
In the present work we have discussed the combined theory of general relativity and scalar QED, 
and observed that it is possible to treat this theory too as an effective field theory, and hence
avoid the traditional renormalization problems. What is more important --- quantum corrections to the theory can be
calculated explicitly, and treated perturbatively as in any other quantum field theory. 

Certainly the effective field theory approach is only valid at sufficiently low energies, 
i.e. below the Planck scale $\sim 10^{19}\ {\rm GeV}$, and at long distances. At higher energies
a new unknown theory will have to govern the quantum gravitational effects. Perhaps 
some kind of string theory, compactified at low energies. However the Planck energy
scale is much larger than 'traditional' high energy scales. Similarly the standard model
may only be a good description below $\sim 1000 \ {\rm GeV}$, so the effective field theory
approach is seemingly good for all energies we are presently dealing with in high energy
physics.  

However experimentally it will be very difficult to verify a quantum gravitational theory ---
even in the presence of charged scalars.

The non-relativistic potential may not be the best offset for a verification of quantum
gravity. This is because the potential has coordinate dependent terms. 
It may be of more interest to look for definite coordinate system invariant
expectations --- in the quest for an experimental verification of quantum gravity. 

\begin{acknowledgments}
I would like to thank P.H. Damgaard for many interesting discussions and useful comments.
\end{acknowledgments}

\appendix
\section{Summary of the vertex rules}
\subsubsection{Scalar propagator}\noindent
The massive scalar propagator is well known:\\ \\

\begin{minipage}[h]{0.4\linewidth}\vspace{0.4cm}
\centering\includegraphics[scale=1.1]{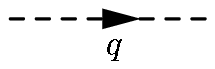}
\end{minipage}
\begin{minipage}[h]{0.65\linewidth}
\centering$\displaystyle = \frac i{q^2-m^2+i\epsilon}$
\end{minipage}

\subsubsection{Photon propagator}\noindent
The photon propagator is also known from the literature. 
We have applied Feynman gauge which gives the least 
complicated propagator:\\ \\

\begin{minipage}[h]{0.4\linewidth}\vspace{0.4cm}
\centering\includegraphics[scale=1.1]{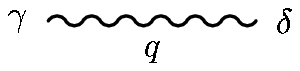}
\end{minipage}
\begin{minipage}[h]{0.65\linewidth}
\centering $\displaystyle = \frac {-i\eta^{\gamma\delta}}{q^2+i\epsilon}$
\end{minipage}

\subsubsection{Graviton propagator}\noindent
The graviton propagator in harmonic gauge is discussed in the literature~\cite{Veltman1,Donoghue:dn}, 
but can be derived quite easily explicitly~\cite{Thesis}. We shall write it in the form:\\ \\

\begin{minipage}[h]{0.4\linewidth}\vspace{0.4cm}
\centering\includegraphics[scale=1.1]{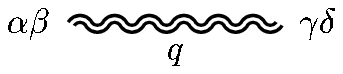}
\end{minipage}
\begin{minipage}[h]{0.65\linewidth}
\centering $\displaystyle = \frac {i{\cal P}^{\alpha\beta\gamma\delta}}{q^2+i\epsilon}$
\end{minipage}\\ 
where $${\cal P}^{\alpha\beta\gamma\delta} = \frac12\left[\eta^{\alpha\gamma}\eta^{\beta\delta} +
\eta^{\beta\gamma}\eta^{\alpha\delta}
-\eta^{\alpha\beta}\eta^{\gamma\delta}\right]$$

\subsubsection{2-scalar-1-photon vertex}\noindent
The 2-scalar-1-photon vertex is well known in the literature. We will
write this vertex as:\\

\begin{minipage}[h]{0.4\linewidth}\vspace{0.15cm}
\centering\includegraphics[scale=1.3]{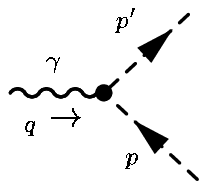}
\end{minipage}
\begin{minipage}[h]{0.65\linewidth}
\centering$\displaystyle = \tau_1^{\gamma}(p,p',e)$
\end{minipage}\vspace{0.3cm}
where 
$$\tau_1^{\gamma}(p,p',e) = -ie\left(p+p'\right)^\gamma$$

\subsubsection{2-scalar-1-graviton vertex}\noindent
The 2-scalar-1-graviton vertex is also discussed in the literature~\cite{Donoghue:dn,Thesis}. 
We will write it in the following way:\\

\begin{minipage}[h]{0.4\linewidth}\vspace{0.15cm}
\centering\includegraphics[scale=1.3]{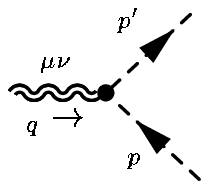}
\end{minipage}
\begin{minipage}[h]{0.65\linewidth}
\centering$\displaystyle = \tau_2^{\mu \nu}(p,p',m)$
\end{minipage}\vspace{0.3cm}
where 
$$\tau_2^{\mu\nu}(p,p',m) = -\frac{i\kappa}2\left[p^\mu p^{\prime \nu}
+p^\nu p^{\prime \mu} - \eta^{\mu\nu}\left((p\cdot p^\prime)-m^2\right)\right]
$$

\subsubsection{2-photon-1-graviton vertex}\noindent
For the 2-photon-1-graviton vertex we have derived:\\

\begin{minipage}[h]{0.4\linewidth}\vspace{0.15cm}
\centering\includegraphics[scale=1.3]{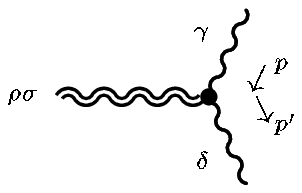}
\end{minipage}
\begin{minipage}[h]{0.65\linewidth}
\centering$\displaystyle = \tau_3^{\rho\sigma(\gamma \delta) }(p,p')$
\end{minipage}\vspace{0.3cm}
where 
\begin{widetext}\nonumber
$$\tau_3^{\mu\nu(\gamma\delta)}(p,p') = i\kappa\Big[{\cal P}^{\rho\sigma(\gamma\delta)}(p\cdot
p') +\frac12\Big(\eta^{\rho\sigma} p^\delta p^{\prime\gamma} + \eta^{\gamma\delta}
(p^\rho p^{\prime \sigma} + p^\sigma p^{\prime \rho})- (p^{\prime\gamma}p^\sigma
\eta^{\rho\delta} + p^{\prime\gamma}p^\rho
\eta^{\sigma\delta} + p^{\prime\rho}p^\delta
\eta^{\sigma\gamma} + p^{\prime\sigma}p^\delta
\eta^{\rho\gamma})\Big)\Big]$$
\end{widetext}

\subsubsection{2-scalar-2-photon vertex}\noindent
The 2-scalar-2-photon vertex is also well known from scalar QED. We will
write it as:\\

\begin{minipage}[h]{0.4\linewidth}\vspace{0.15cm}
\centering\includegraphics[scale=1.3]{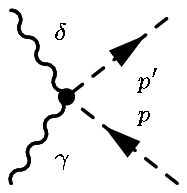}
\end{minipage}
\begin{minipage}[h]{0.65\linewidth}
\centering $\displaystyle = \tau_4^{\gamma\delta}(p,p',e)$
\end{minipage}\vspace{0.3cm}
where
$$\tau_4^{\gamma\delta}(p,p',e) = 2ie^2\eta^{\gamma\delta}$$

\subsubsection{2-scalar-1-photon-1-graviton vertex}\noindent
For the 2-scalar-1-photon-1-graviton vertex we have derived:\\

\begin{minipage}[h]{0.4\linewidth}\vspace{0.15cm}
\centering\includegraphics[scale=1.3]{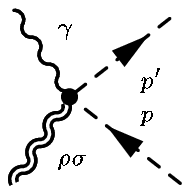}
\end{minipage}
\begin{minipage}[h]{0.65\linewidth}
\centering $\displaystyle = \tau_5^{\rho\sigma(\gamma)}(p,p',e)$
\end{minipage}\vspace{0.3cm}
where
$$\tau_5^{\rho\sigma(\gamma)}(p,p',e) = ie\kappa\left[{\cal P}^{\rho\sigma\alpha\gamma}(p+p')_\alpha\right]$$
and ${\cal P}^{\rho\sigma\alpha\gamma}$ is defined as above.

For all vertices the rules of momentum conservation has been applied. 
For the external scalar lines we associate a factor of 1. At each loop we will
integrate over the undetermined loop momentum.

For a certain diagram we will divide with the appropriate symmetry factor
of the Feynman diagram.  

\begin{widetext}
\section{The needed integrals for the calculation of the diagrams}
To calculate the diagrams the following integrals are needed
\begin{eqnarray}\displaystyle
J=&\displaystyle \int\frac{d^4l}{(2\pi)^4} \frac{1}{l^2(l+q)^2} & = \frac{i}{32\pi^2}\big[-2L\big] + \ldots\\
J_\mu=&\displaystyle\int\frac{d^4l}{(2\pi)^4} \frac{l_\mu}{l^2(l+q)^2} & =   \frac{i}{32\pi^2}\Big[q_\mu L\Big]+\ldots\\
J_{\mu\nu}=&\displaystyle\int\frac{d^4l}{(2\pi)^4} \frac{l_\mu l_\nu}{l^2(l+q)^2} 
& = \frac{i}{32\pi^2}\bigg[q_\mu q_\nu \Big(-\frac23L\Big) - q^2\eta_{\mu\nu}\Big(-\frac16 L\Big)\bigg]+\ldots
\end{eqnarray}
together with
\begin{eqnarray}\displaystyle
I=&\displaystyle\int\frac{d^4l}{(2\pi)^4} \frac{1}{l^2(l+q)^2((l+k)^2-m^2)} & = \frac{i}{32\pi^2m^2}\big[-L-S\big]+\ldots\\
I_\mu=&\displaystyle\int\frac{d^4l}{(2\pi)^4} \frac{l_\mu}{l^2(l+q)^2((l+k)^2-m^2)} & =  \frac{i}{32\pi^2m^2}\bigg[k_\mu\bigg(\Big(-1-\frac12 \frac{q^2}{m^2}\Big)L-
\frac14\frac{q^2}{m^2}S\bigg)+q_\mu\bigg(L+\frac12S\bigg)\bigg]+\ldots\\
I_{\mu\nu}=&\displaystyle\int\frac{d^4l}{(2\pi)^4} \frac{l_\mu l_\nu}{l^2(l+q)^2((l+k)^2-m^2)} & =  \frac{i}{32\pi^2m^2}\bigg[q_\mu q_\nu\bigg(-L-\frac38 S\bigg)
+k_\mu k_\nu \bigg(-\frac12\frac{q^2}{m^2}L-\frac18\frac{q^2}{m^2}S\bigg)\\
&&+\big(q_\mu k_\nu + q_\nu k_\mu \big)\bigg(\Big(\frac12 + \frac12\frac{q^2}{m^2}\Big)L + \frac{3}{16}\frac{q^2}{m^2}S\bigg)+
q^2\eta_{\mu\nu}\Big(\frac14L+\frac18S\Big)\nonumber
\bigg]+\ldots 
\end{eqnarray}
\begin{eqnarray}\displaystyle
I_{\mu\nu\alpha}=&\displaystyle\int\frac{d^4l}{(2\pi)^4} \frac{l_\mu l_\nu l_\alpha}{l^2(l+q)^2((l+k)^2-m^2)} & =  \frac{i}{32\pi^2m^2}\bigg[
q_\mu q_\nu q_\alpha\bigg(L+\frac5{16}S\bigg)+k_\mu k_\nu k_\alpha\bigg(-\frac16 \frac{q^2}{m^2}\bigg)
\nonumber\\ \nonumber&&+\big(q_\mu k_\nu k_\alpha + q_\nu k_\mu k_\alpha + q_\alpha k_\mu k_\nu\big)\bigg(\frac13\frac{q^2}{m^2}L+
\frac1{16}\frac{q^2}{m^2}S\bigg)
\\&&+\big(q_\mu q_\nu k_\alpha + q_\mu q_\alpha k_\nu + q_\nu q_\alpha k_\mu \big)\bigg(\Big(-\frac13 - \frac12\frac{q^2}{m^2}\Big)L
-\frac{5}{32}\frac{q^2}{m^2}S\bigg)
\\ \nonumber &&+\big(\eta_{\mu\nu}k_\alpha + \eta_{\mu\alpha}k_\nu + \eta_{\nu\alpha}k_\mu\big)\Big(\frac1{12}q^2L\Big)
\\ \nonumber&&+\big(\eta_{\mu\nu}q_\alpha + \eta_{\mu\alpha}q_\nu + \eta_{\nu\alpha}q_\mu\big)\Big(-\frac16q^2L -\frac1{16}q^2S\Big)
\bigg]+\ldots 
\end{eqnarray}
where $L=\ln(-q^2)$ and $S=\frac{\pi^2m}{\sqrt{-q^2}}$. In the above integrals only the lowest order non-analytical terms are
presented. The ellipses denote higher order non-analytical contributions as well as the neglected analytical terms. Furthermore the 
following identities hold true for the on shell momenta, $k\cdot q = \frac{q^2}2$, where $k-k'=q$ and $k^2=m^2=k^{\prime 2}$. In some cases the integrals
are needed with $k$ replaced by $-k'$, where $k' \cdot q = -\frac{q^2}{2}$, these results, are obtained by replacing everywhere 
$k$ with $-k'$. This can be verified explicitly. The above integrals checks with the results of~\cite{Donoghue:dn}.
  
The following integrals are needed to do the box diagrams. The ellipses denote higher
order contributions of non-analytical terms as well as neglected analytical terms~\cite{Donoghue3}.
\begin{eqnarray}\displaystyle
K  = &\displaystyle \int\frac{d^4l}{(2\pi)^4} \frac{1}{l^2(l+q)^2((l+k_1)^2-m_1^2)((l-k_3)^2-m_2^2)} = &  \frac{i}{16\pi^2m_1 m_2 q^2}\bigg[\Big(1-\frac{w}{3m_1m_2}\Big)L\bigg]+\ldots \\
K' = &\displaystyle \int\frac{d^4l}{(2\pi)^4} \frac{1}{l^2(l+q)^2((l+k_1)^2-m_1^2)((l+k_4)^2-m_2^2)} = &  \frac{i}{16\pi^2m_1 m_2 q^2}\bigg[\Big(-1+\frac{W}{3m_1m_2}\Big)L\bigg]+\ldots
\end{eqnarray}
Here $k_1\cdot q = \frac{q^2}2$, $k_2\cdot q = -\frac{q^2}2$, $k_3\cdot q = -\frac{q^2}2$ and $k_4\cdot q = \frac{q^2}2$, where $k_1-k_2=k_4-k_3=q$ and $k_1^2=m_1^2=k_2^2$ 
together with $k_3^2=m_2^2=k_4^2$. Furthermore we have defined $w = (k_1\cdot k_3)-m_1m_2$ and $W = (k_1\cdot k_4) - m_1m_2$. The above results for the integrals 
checks with~\cite{Donoghue:1996mt}.   

For the above integrals the following constraints for the non-analytical terms can be verified directly on
the mass-shell:
\begin{equation}
I_{\mu\nu\alpha}\eta^{\alpha\beta} = I_{\mu\nu}\eta^{\mu\nu} = J_{\mu\nu}\eta^{\mu\nu} = 0
\end{equation}
\begin{equation} 
I_{\mu\nu\alpha}q^\alpha = -\frac{q^2}{2}I_{\mu\nu}, \ \ \ \ I_{\mu\nu}q^\nu  = -\frac{q^2}{2}I_{\mu}, \ \ \ \   I_{\mu}q^\mu  = -\frac{q^2}{2}I \ \ \ \
J_{\mu\nu}q^\nu = -\frac{q^2}{2}J_{\mu}, \ \ \ \ J_{\mu}q^\mu   = -\frac{q^2}{2}J  
\end{equation}
\begin{equation}
I_{\mu\nu\alpha}k^\alpha = \frac{1}{2}J_{\mu\nu}, \ \ \ \ I_{\mu\nu}k^\nu  = \frac{1}{2}J_{\mu}, \ \ \ \   I_{\mu}k^\mu  = \frac{1}{2}J  
\end{equation}
These mass-shell constraints can be used to derive the above integrals and are directly verified in the same manner, that we simplify the $K$ and $K'$ integrals in the 
box calculations. 
\end{widetext}


\begin{thebibliography}{99}
\bibitem{Iwasaki2}
{For an alternative definition of the potential, see \cite{Iwasaki} where the second 
Born term is taken into account.}
\bibitem{Donoghue3}{We have not written the imaginary parts of the $K$-integral explicitly down and have neglected
them in the calculations. It is believed~\cite{Donoghue:1996mt} that these parts exponentiate to a gravitational 
equivalent of the Coulomb phase which has no importance for the squared amplitude.}
\end{thebibliography}
\end{document}